\documentclass[12pt]{article}
\newcommand{\be}{\begin{equation}}
\newcommand{\ee}{\end{equation}}
\newcommand{\beas}{\begin{eqnarray*}}
\newcommand{\eeas}{\end{eqnarray*}}
\newcommand{\bea}{\begin{eqnarray}}
\newcommand{\eea}{\end{eqnarray}}
\newcommand{\ba}{\begin{array}}
\newcommand{\ea}{\end{array}}
\newcommand{\nn}{\nonumber}

\newcommand{\e}{\epsilon}

\newcommand{\g}{\gamma}
\newcommand{\de}{\delta}
\newcommand{\si}{\sigma} 

\newcommand{\dal}{{\sqcap\!\!\!\!\sqcup}}
\begin{document}
\title{
{\normalsize
\begin{flushright}
{IHEP 2005-15
}\\
\end{flushright} }
\vspace{2ex}
{\bf General covariance violation  and\\ 
the gravitational dark matter.\\
I. Scalar graviton}}
\author{Yu.\ F.\ Pirogov
\\[0.5ex]
{\it Theory Division, Institute for High Energy Physics,  Protvino,}\\
{\it  RU-142281 Moscow Region, Russia}
}
\date{}
\maketitle
\abstract{\noindent 
The violation of the general covariance is proposed as a 
resource of the gravitational dark  matter.
The minimal violation of the covariance to the 
unimodular one is associated with  
the massive scalar graviton  as the simplest representative of such a
matter. The Lagrangian formalism for the continuous medium, the
perfect fluid in particular,
in the scalar graviton environment is developed. 
The implications for cosmology are shortly indicated.
} 

\section{Introduction}

The General Relativity (GR) 
is the viable theory of  gravity, perfectly consistent   with  
observations. Nevertheless, it encounters a number of 
the conceptual problems among which one can mention that of the dark
matter (DM). To solve the
latter problem one adjusts usually the conventional  or new matter
particles, remaining still in the realm of the GR (for a recent review
see, e.g., ref.~\cite{DM}).  
The ultimate goal of the DM being in essence to participate
only in the gravitational interactions,
one can try and attribute  to this purpose the gravity itself, going
thus  beyond the GR. 

Namely, the GR is  the field theory of the massless
tensor graviton as a part of the metric.  However, the metric 
contains also the extra degrees of freedom which could be associated
with  the massive scalar, vector and  scalar-tensor gravitons. 
Nevertheless, the latter ones lack the explicit physics
manifestations. 
This is due to the fact that the GR incorporates as the basic
symmetry the general covariance~(GC). The latter serves as the gauge
symmetry to remove from the observables the degrees of freedom in
excess of the massless tensor graviton.
So, to associate the extra particles contained in
the metric with  (a part of) the DM the violation of the GC
is obligatory. 

Starting from the GR, an arbitrary variety of the GC violations is
admissible. Contrary to this, a hierarchy of the GC
violations is expected
in the framework of the affine Goldstone approach to
gravity~\cite{Pirogov1}. In distinction with the GR, the latter
approach  is based on two symmetries:   
the global affine symmetry (AS) plus the GC. As a result, there are
conceivable two types of the GC violations: those with and without the
explicit AS violation.
The first type of violation  is to be strongly suppressed.
But the second one is a priori arbitrary. 
In the reasonable assumptions, the GC  violation can be
parametrized in the given approach through the background
metric.\footnote{The basic ingredient of the approach is  the
spontaneous appearence of the metric in the affinely connected world.
For the short exposition, see ref.~\cite{Pirogov2}. 
For a related development, see ref.~\cite{Carmeli}. } 

Marginally, the latter description  can depend  only on
the determinant of the background
metric. In this case, there still resides the unimodular covariance
(UC), i.e., that relative to the volume preserving diffeomorphisms.
In comparison with the GC, the UC lacks only the local scale
transformations.
Being  next-of-kin to the GC, the UC suffices  to incorporate
both the massive scalar graviton and the massless
tensor one, but nothing redundant.\footnote{In fact, it is the UC, not
the GC, which ensures the masslessness of the tensor
graviton~\cite{Vanderbij}. In other contexts, the UC is
studied in refs.~\cite{UC}~--~\cite{Tiwari}.} The
violation of the GC to the residual UC is considered in the present
paper to be the raison d'etre for the existence of
the scalar graviton as a simplest representative of the
gravitational DM.
More elaborate dependencies on the background metric, with the
residual covariance group being even more restricted, could correspond
to the massive vector and  scalar-tensor gravitons.

In the present paper, we confine ourselves by the scalar graviton,
the vector and scalar-tensor ones being postponed to the future.
In Sec.~2,  the GC violation, in general,  and, specifically,  the
scalar graviton are considered. The Lagrangian description of the
continuous medium, the perfect fluid in particular, in the scalar
graviton environment is developed in Sec.~3, with some remarks in
Conclusion. 

\section{Gravity}
\paragraph{Gravitational DM}

Consider the classical theory of the  metric $g_{\mu\nu}$ and
the generic  matter field $\phi$ with the action 
\be\label{GCV}
I=\int\Big(L_{g}(g_{\mu\nu})+ \Delta L_{g}(g_{\mu\nu}, \bar \kappa_p)
+ L_m( \phi, \bar \kappa_p) \Big)
\sqrt{-g}\,d^4x.
\ee
Here $L_{g}$ and  $\Delta L_g$ are, respectively,  the generally
covariant  and  GC violating contributions of the gravity,
$L_m$ being the matter Lagrangian. In general,  the latter violates
the GC, too. All the Lagrangians above are assumed to be the scalars. 
The functions $\bar \kappa_p=\bar \kappa_p(x^\mu)$,
$p=1,\dots$, with $\bar \kappa'_p(x'^{\mu})\neq\bar
\kappa_p(x^\mu)$ parametrize the GC violation.\footnote{We
do not consider the  space-time dependence of the  scalar
parameters, such as the Newton's constant, etc.}    The number and
the transformation properties of these background parameter-functions
relative  to the change of the coordinates~$x^\mu\to x'^{\mu}$
determine ultimately the group of the residual covariance. The
parameter-functions
are the generalization of the constant parameters and are to be found
through observations.  The space-time dependence of $\bar\kappa_p$
tacitly implies the existence of the distinguished
coordinates relative to some background. 
This means that the metric Universe, contrary to what is assumed in
the GR, is not a self-contained system and  can not exclusively be
described in the internal dynamical terms.
Conceivably, the background parameter-functions will also be
put on the dynamical footing in the course of time.

It is well known, that any generally covariant Lagrangian $L_g$
corresponds to the massless tensor graviton 
(see, e.g., ref.~\cite{Vanderbij}). The
reason is that the GC removes in the  equations of motion all
the degrees of freedom contained in the metric but those with the
helicities $\lambda=\pm 2$. To make the  rest of the metric
degrees of freedom  to be observable, the addition of the GC violating
$\Delta L_g$ is obligatory.

Varying the action~(\ref{GCV}) with respect to $g^{\mu\nu}$ 
one arrives at the gravity equation of motion:
\be\label{eomg}
G_{\mu\nu}+ \Delta G_{\mu\nu} = M_P^{-2}\, T^{(m)}_{\mu\nu}.
\ee
Here   $G_{\mu\nu}$ is the gravity tensor defined as follows:
\be\label{G}
-M_P^{2} G_{\mu\nu}
=\frac{2}{\sqrt{-g}}\,
\frac{\hspace{-1ex}\de {\cal L}_g}{\de
g^{\mu\nu}}
=2\,\frac{\hspace{-1ex}\de  L_g}{\de g^{\mu\nu}}-
L_g g_{\mu\nu}
\ee
with ${\cal L}_g={\sqrt{-g}}\, L_g$ being the Lagrangian density 
(and similarly for $\Delta G_{\mu\nu}$). 
In the above, one puts conventionally  $M_P^{-2}=8\pi G_N$, with $M_P$
being the Planck mass and $G_N$ being the Newton's constant.
$T^{(m)}_{\mu\nu}$ is the conventional energy-momentum tensor of
the matter defined  through  $L_m$ by the r.h.s.\ of eq.~(\ref{G}).
 
Due to the general covariance of $L_g$, one arrives at the identity:
\be
\nabla_\nu G^{\mu\nu}=0,
\ee
with $\nabla_\nu$ being the covariant derivative 
defined by the metric $g_{\mu\nu}$.
Thus with account for  eq.~(\ref{eomg}), the  modified conservation
condition  
\be\label{cc}
\nabla_\nu (T_m^{\mu\nu} -M_P^{2} \Delta G^{\mu\nu})=0
\ee
is to be fulfilled. Clearly, the energy-momentum of the ordinary
matter alone ceases to conserve. To restore the conventional
interpretation of the conserved matter, one
should consider the extra gravity degrees of freedom  as (a part of)
the DM with the energy-momentum tensor
\be
\Delta T_{g}^{\mu\nu}= -M_P^{2} \Delta G^{\mu\nu}.
\ee
The respective particles have no specific
quantum numbers and  possess only the
gravitational interactions, so that such an association is quite
natural. 

In other terms, eq.~(\ref{cc}) can  be written~as
\be\label{cl1}
\nabla_\nu T_m^{\mu\nu}=Q^\mu,
\ee
with the vector  
\be\label{F}
Q^\mu=M_P^{2}\nabla_\nu \Delta G^{\mu\nu}
\ee
representing the extra gravity force acting on the ordinary matter.
This force reflects the nonfulfilment of the equivalence principle
due to the GC violation. 
In the case that $L_m$ does not depend on
$\bar \kappa_p$  and  thus is generally covariant,  one gets with
account for the matter equation of motion: 
\be\label{00}
\nabla_\nu T_m^{\mu\nu}= Q^\mu=0.
\ee

\paragraph{Scalar graviton}

In what follows, we restrict ourselves by the  minimal violation  of
the GC with one parameter-function,  the scalar density, and
respectively  with the scalar graviton alone as the simplest
representative of the gravitational DM.
Conventionally, take as~$L_g$  the modified 
Einstein-Hilbert Lagrangian:
\be\label{EH}
L_g=- M_P^2 \Big(\frac{1}{2}R(g_{\mu\nu})-\Lambda\Big),
\ee
with $R$ being the Ricci scalar and $\Lambda$ being the
cosmological constant (if any). Eq.~(\ref{EH})
describes the massless tensor graviton. 

To incorporate additionally only the scalar graviton, the minimal GC
violation with the residual UC is sufficient.  
To this end, use  as the respective field variable the relative metric
scale $e^{\chi} =\sqrt {-g}/\sqrt{-\bar g}$ or, otherwise, 
\be
\chi=\frac{1}{2}\ln \frac{g}{\bar g}.
\ee
Here one puts $g=\det g_{\mu\nu}$ and $\bar g=\det \bar g_{\mu\nu}$,
with $g_{\mu\nu}$ and $\bar g_{\mu\nu}$ being  the  dynamical and
background metrics, respectively. Being
determined by the ratio of two scalar densities, $\chi$ itself
is the scalar. In other terms, it can be  expressed  as follows:
\be
\partial_\mu \chi=\Gamma^\lambda{}_{\lambda\mu}-
\bar\Gamma^\lambda{}_{\lambda\mu},
\ee
with $\Gamma^\lambda{}_{\nu\mu}$ and  $\bar\Gamma^\lambda{}_{\nu\mu}$
being, respectively,  the dynamical and background Christoffel
connections. Clearly, the difference of the two transforms
homogeneously under the change of the coordinates. 

For the physics clarity, fix a coordinate system
and put in this system by definition
$g_{\mu\nu}\equiv (-\bar  g)^{1/4}
(\eta_{\mu\nu}+h_{\mu\nu})$, with
$\eta_{\mu\nu}$ being the Minkowski tensor. Then one gets in the
linearized approximation $\chi\simeq 1/2\, \eta^{\mu\nu}h_{\mu\nu}$,
in accord with treating~$\chi$ as the scalar
graviton. Note that knowing $\bar g$ in the observer's coordinates
$x^\mu$, one can always find the background-attached
coordinates $\bar x^\mu$, where $\bar g=-1$.

Take the scalar graviton Lagrangian in the simplest form:
\be\label{Ls}
\Delta L_g=\frac{1}{2}\mu^2\,\partial
\chi\cdot\partial \chi -V(\chi).
\ee
Here and in what follows, the notation $A\cdot B=A_\mu B^\mu$ is used
for any two vectors.
In the above, $\mu$ is a constant with the dimension of
mass.\footnote{Were it not for $\mu\neq 0$, $\chi$ would become the
auxiliary nonpropagating field.  A priori, the mass squared of the
scalar graviton may be negative, with $\chi$ corresponding to the
tachion instead of the physical particle.  One could  envisage also
$\mu^2< 0$, with $\chi$ corresponding to the ghost.} 
The variable $\mu\chi$ is the conventionally normalized field of
the scalar graviton. 

Due to dependence on $\bar g$, the GC
is violated. Nevertheless, the Lagrangian
possesses the residual UC, i.e., that relative
to the variations   of  the coordinates $\delta x^\mu=\epsilon^\mu$
with the auxiliary condition $\partial\cdot \epsilon=0$. This
leaves~$\bar g$ (in the line with~$g$)  unchanged.
The derivative part of the Lagrangian above does not violate the
AS~\cite{Pirogov1}, hence  the constant $\mu$ is not a priori
suppressed relative to~$M_P$. The potential $V(\chi)$ ascribes  mass
to the scalar graviton. Violating the AS~\cite{Pirogov1}, the
$\chi$-dependent part of the potential is to be suppressed. As for the
$\chi$-independent part, one puts conveniently 
$V(0)=0$ attributing the constant part to the $\Lambda$-term.

As for the matter field $\phi$,  the  generally
covariant Lagrangian $L_m(D_\mu\phi, \phi)$, with $D_\mu\phi$ being
the generally covariant derivative, should be substituted by the
Lagrangian $L_m(D_\mu\phi, \phi, \partial_\mu\chi, \chi)$. Similarly
to the gravity Lagrangian, the
derivative  $\chi$-terms do not violate the AS, whereas
those without derivative do violate it and thus should be suppressed. 

Varying the total action 
with respect to $g^{\mu\nu}$ ($\bar g$ being unchanged under the
variation) and taking into
account the relations $\delta g_{\lambda\rho}=-
g_{\lambda\mu}g_{\rho\nu}\,\delta g^{\mu\nu}$,  $\delta\sqrt{-g}
=-1/2\,\sqrt{-g}g_{\mu\nu}\delta g^{\mu\nu}$, so that
$\delta \chi=-1/2\, g_{\mu\nu}\delta g^{\mu\nu}$,
one arrives at the equation of motion for gravity~(\ref{eomg}),
where 
\be
G_{\mu\nu}=R_{\mu\nu}-\frac{1}{2}R\,g_{\mu\nu}+\Lambda
g_{\mu\nu}
\ee
and 
\be\label{sg}
\Delta G_{\mu\nu}=-M_P^{-2}\Big(\mu^2 (\partial_\mu
\chi\partial_\nu \chi
-\frac{1}{2}\,\partial
\chi\cdot\partial \chi\, g_{\mu\nu}+\dal \chi\, g_{\mu\nu})+
(V +\partial V/ \partial \chi) g_{\mu\nu}\Big),
\ee
with $\dal$ being the covariant Dalambertian: 
\be\label{dal}
\dal\equiv
\nabla\cdot \nabla=\frac{1}{\sqrt{-g}}\,\partial\cdot
(\sqrt {-g}\,  \partial).
\ee
Thereof one gets for the scalar gravity force,  eq.~(\ref{F}),
\be
Q_\mu=- \mu^2(\dal \chi\,\partial_\mu
\chi
+\partial_\mu \dal\chi) -  \partial_\mu(V +\partial V/ \partial
\chi).
\ee
From the gravity equation
of motion~(\ref{eomg}), there follows for the scalar curvature:
\be
R=4\Lambda+M^{-2}_P\Big(\mu^2(\partial
\chi\cdot\partial \chi-4\dal \chi)
-4(V+\partial V/ \partial \chi)-T_m{}^\mu_\mu\Big).
\ee
Contrary to the GR, $R$ can be nonzero even in the absence of
the $\Lambda$-term  and  matter. Conceivably, the gravitational
DM alone, the scalar graviton in particular,  suffices to
form the (quasi)-stable gravitational configurations.

\section{Continuous medium}

\paragraph{Distorted medium}

Having in mind the application to cosmology, choose  as the matter the
continuous medium.  
Such a macroscopic approach is to be applicable at the 
sufficiently long length  scales compared to the mean free path of the
matter particles, with the quantum coherence effects being averaged.
Conventionally, one restricts himself by the
description of the continuous medium at the level of the equations of
motion directly in terms of the energy-momentum tensor. This, in
principle, suffices under the validity of the GC. 
To comprehend the new effects produced  the scalar graviton one has
to adhere to the Lagrangian framework. 

The mathematically strict Lagrangian description of the  continuous
medium proceeds in terms of the comoving coordinates $z^a$ and
$z^a_\mu\equiv \partial_\mu z^a$, $a=0,\dots, 3$, as the Lagrangian 
variables (see, e.g., ref.~\cite{Jacques}). $z^0$ is closely related
with the proper time for a medium element, so that 
$\partial/\partial z^0 \sim U\cdot\partial$, 
with $U^\mu$ being the medium 4-velocity. 
$z^a$, $a=1,2,3$, are the fields which label the streamlines. 
Using these fields, one  can
automatically account for the continuity condition, the latter being
the generic  property of the continuous media. Inasmuch as such a
consistent description is possible, we develop in what
follows the  more straightforward formalism directly in terms of the
observer's $x^\mu$ coordinates.

To describe the medium at the Lagrangian level, we choose, in addition
to $U^\mu$, $U\cdot U=1$, the  proper (i.e.,  measured in the
comoving coordinates)  concentration $n$ of the medium particles 
and  the specific entropy~$\si$  (the entropy per particle).  
Equivalently, one could chose as the pair of the independent dynamical
variables the entropy $s=n\si$ per unit proper volume and $\si$.
Besides, the medium is characterized by the nondynamical
parameters such as the particle mass~$m$, etc.  
The medium has to satisfy the continuity condition for the particle 
number current $N^\mu=n U^\mu$
\be\label{nN}
\nabla\cdot N=\frac{1}{\sqrt{-g}}\,
\partial\cdot (\sqrt{-g}N)=0,
\ee 
This constraint has to be  valid identically, independently of the
equations of motion. 

Take the Lagrangian for the continuous medium generically as
\be\label{P}
L_m(U^\mu\!,n,\si)= - E (|N|,\si),
\ee
with $E (|N|,\si)$ being the  scalar energy function. Here
one puts $|N|= (N\cdot N)^{1/2}$.
One can also add  the derivative interactions of the scalar
graviton with the medium  as follows:
\be
L'_{m}(U^\mu\!,n,\si)=-  F(|N|,\si)\frac{1}{|N|}\,  N\cdot
\partial \chi,
\ee
with $F(|N|,\si)$ being the scalar formfactor. Violating the
AS, the derivativeless $\chi$-dependence is supposed to be
absent.\footnote{The terms quadratic in $\partial_\mu \chi$, being
treated as the corrections to the Lagrangian~$\Delta L_g$, are
neglected in the first approximation.}

The continuity condition (\ref{nN}) is clearly equivalent to the
ordinary conservation of the vector density
${\cal N}^\mu=\sqrt{-g}\,N^\mu$, 
$\partial\cdot {\cal N} =0$,
which does not include the metric. Not to vary the continuity
condition under the change of the metric, it is 
${\cal N}^\mu$, not $N^\mu$,  to be taken as the
independent variable.
Introducing  the scalar density $|{\cal N}|=({\cal N} \cdot
{\cal N})^{1/2}=\sqrt{-g}\,| N|$, wright finally  the  
total effective Lagrangian density  as follows:
\be\label{TLd}
{\cal  L}_m^{(tot)}  =
-\sqrt{-g}\bigg( E \Big(\frac{|{\cal N}|}{\sqrt{-g}}\,,
\si\Big)+F \Big(\frac{|{\cal N}|}{\sqrt{-g}}\, ,
\si\Big) \frac{1}{|{\cal N}|}\, {\cal N}\cdot\partial \chi\bigg)
+\lambda\,
\partial\cdot {\cal N}. 
\ee
where  $\lambda$ is the Lagrange's multiplier.

Varying eq.~(\ref{TLd}) with respect to $\lambda$ one
reproduces the constraint eq.~(\ref{nN}). Further, varying  
${\cal L}_m^{(tot)}$
relative to $g^{\mu\nu}$ one gets the energy-momentum tensor: 
\be\label{T}
T^{(m)}_{\mu\nu}=(\rho+p)
U_\mu U_\nu- p\, g_{\mu\nu}, 
\ee
with
\bea\label{rhop}
\rho&=&e -\nabla\cdot(f U),\nn\\
p&=&  n e'-e +(n f'-f)  U\cdot\partial \chi 
+\nabla\cdot(f U),
\eea
where $e (n,\si)\equiv E (|N|,\si)\vert_{|N|=n}$, $e'=\partial e
(n,\si)/\partial n$ 
and similarly for $f$. The trace of the energy-momentum
tensor is as usually
\be
T_{m}{}^\mu_\mu=\rho-3p.
\ee
In the above, $\rho$ is the scalar
coinciding with the energy  per unit proper 
volume,  $p$ is the scalar coinciding with the (isotropic) pressure,
while  $f $ is the new scalar state function. 
The terms proportional to~$f$ distort the medium. Being of the odd
degree in velocity these terms reflect the dissipation (pumping)
caused by the environment of the scalar gravitons. 

The dynamical variables $U^\mu$, $n$ and $\si$ have to satisfy the
modified  conservation condition
eq.~(\ref{cc}) and  the continuity condition eq.~(\ref{nN}).
This gives five equations for the same number
of the independent variables. Supplemented by the ten gravity
equations of motion~(\ref{eomg}) for the metric $g_{\mu\nu}$ and
thus~$\chi$,  this   presents the complete  system of
the dynamical equations. 

Projecting eq.~(\ref{cc}) on the streamlines by  multiplying on
$U_\mu$ and accounting for $U\cdot \nabla_\nu U=0$, one gets the
energy equation in the external field: 
\be\label{ee}
\nabla\cdot\Big((\rho+p )U\Big)-
U\cdot\partial p = U\cdot Q.
\ee
The scalar $q=U\cdot Q$
on the r.h.s.\ of the equation above is nothing but the
power $Q_0$ dissipating from (depositing in)  the medium per unit
proper volume.
Likewise, restricting eq.~(\ref{cl1}) by the projector 
$P^{\mu\nu}=g^{\mu\nu}-U^\mu U^\nu$,  $P^{\mu\nu}U_\nu=0$,  on the
hypersurface orthogonal to the streamlines, one gets the modified 
Euler equation:
\be\label{Ee}
(\rho+p)U\cdot\nabla U_\mu
+U\cdot\partial p\, U_\mu -\partial_\mu p 
=Q_\mu-  U \cdot Q\, U_\mu.
\ee 
When all the terms but the one proportional to $\rho$ are missing,
eq.~(\ref{Ee}) is nothing but the geodesic condition: $U\cdot\nabla
U_\mu=0$. Otherwise, it describes the deviation of the flow from the
geodesics due to the pressure and the influence of the scalar
gravitons.

\paragraph{Undistorted medium}

Restrict ourselves in what follows by the undistorted medium, $F=0$. 
In this case due to the GC of $L_m$, there should be
fulfilled eq.~(\ref{00}).
The medium is described as usually by the two state functions,
$\rho$ and $p$, depending only on $n$ and $\si$. According
to the second thermodynamic law one has now
\be\label{entropy}
kTd(sv)= d(\rho v)+p\, d v,
\ee
the equality sign referring to the reversible  processes.
In the above, $v$ is the proper volume element, $T$
is the temperature and $k$ is the Boltzmann constant. In particular,
applied to  the one-particle volume, $1/n$, eq.~(\ref{entropy}) reads
\be\label{entropy1}
kTd\si=d\frac{\rho}{n}+p\,d \frac{1}{n},
\ee
which means
\bea\label{p}
p&=&n\frac{\partial\rho}{\partial n}-\rho,\nn\\
kT&=&\frac{1}{n}\frac{\partial\rho}{\partial\si}.
\eea
The first part of the equation above coincides with eq.~(\ref{rhop})
at $f=0$, the second part
being nothing but the thermodynamic definition of the temperature. 

From eq.~(\ref{entropy}), one gets otherwise
\be
kT\frac{d (sv)}{\!\!\!\!\!\!d \tau}= \frac{d((\rho+p)
v)}{\hspace*{-8ex}d \tau} -v\,
\frac{d p}{d \tau}\,,
\ee
with $\tau$ being the proper time. Thereof, with account for the
relations (see, e.g., ref.~\cite{Misner}) 
\bea
\frac{d v}{d \tau}&=&v \nabla\cdot U,\nn\\
\frac{d u}{d \tau}&=& U\cdot\partial u,
\eea
valid for the proper volume $v$ and an arbitrary scalar $u$, one
presents the energy equation~(\ref{ee}) as that for the entropy
conservation:
\be\label{S}
\nabla\cdot S = 0
\ee
with $S^\mu=\si N^\mu$ being the entropy current.
This reflects the adiabaticity of the flow in the absence of the
external force $Q^\mu$.  
With account for the continuity condition, eq.~(\ref{S}) expresses the
isentropy  along the streamlines:
\be\label{si}
U\cdot \partial \si =0.
\ee
Eqs.~(\ref{nN}), (\ref{Ee}) and (\ref{si}), augmented by the equations
of state $\rho=\rho(n,\si)$ and $p=p(n,\si)$,  present the
five equations of motion for the five dynamical variables $U^\mu$, $n$
and $\si$.

The preceding formalism can  be extended to the  multicomponent medium
consisting of the fractions $i=1,\dots, I$.  
In this case, the medium depends generally on the
$(2I+3)$, $I\ge 1$, dynamical variables $U^\mu$, $n_i$ and $\si_i$.
Each one of the medium components 
has to satisfy the continuity condition eq.~(\ref{nN}) for the partial
number currents $N_i^\mu=n_i U^\mu$. For the reversible processes,
one can impose additionally 
the $(I-1)$ thermodynamic equilibrium conditions expressing the
equality of all the partial temperatures $T_i$  to the common
temperature~$T$:
\be
T_i=T.
\ee
The temperatures $T_i$ are to be defined by the counterpart of
eq.~(\ref{p}). The metric $g_{\mu\nu}$ as well as  $\chi$  being
given, this, together with the four conservation equations (\ref{00}),
constitutes $(2I+3)$, $I\ge 1$ equations for the same
number of the independent variables.

\paragraph{Perfect fluid}

Consider one of the medium components, index $i$ being
omitted. The  function $e (s,\si)$  being
given,  one derives by eq.~(\ref{rhop}) the equations
of state $\rho=\rho(n,\si)$ and  $p=p(n,\si)$.
Thereof, one can get the  equation of state  $p=p(n,\rho)$
to be used in cosmology.
And v.v., imposing on the physical ground the last equation and some
thermodynamic relations one can, generally, restore
the respective function $e (s,\si)$. 

To this end,
impose the equation of state in the conventional form of the
proportionality between the pressure and the excitation energy
per the unit proper volume:
\be\label{prho}
p=w(\rho-mn),
\ee
with $w$ being a constant. Put without loss of generality
\be\label{E}
e=(m+\e)n,
\ee
with  $\e(n,\si)$ being a function to be found.
With account for eq.~(\ref{p}), this means
\bea\label{pepsil}
p&=&n^2 \frac{\partial \e}{\partial n},\nn\\
kT&=& \frac{\partial \e}{\partial \si}.
\eea
As the general solution to the  differential
equation ensuing from eq.~(\ref{prho}), one gets
\be\label{eos}
\e= \varepsilon n^{w},
\ee
with $\varepsilon=\varepsilon(\si)$ being an integration constant.

Identifying conventionally for the perfect fluid
\bea\label{rkT}
\rho&=&(m +c_v kT)n,\nn\\
p&=&kTn,
\eea
with $c_v$ being the specific heat, one gets 
\be\label{kT}
kT=w\varepsilon n^{w}, \hspace{1ex} w=c_v^{-1}.
\ee 
On the other hand, from the second part of eq.~(\ref{pepsil}) 
there follows 
\be
kT=
\frac{\partial \varepsilon}{\partial\si} n^{w}.
\ee
Combining the last two expression one gets
\be
\varepsilon=Ce^{w \si}, 
\ee
with $C$ being an integration  constant.
To find the latter constant, one should match, at a critical
temperature $T_c$, the relations obtained for the fluid
with those for the crystal phase setting in the vicinity of $T=0$.
This should  ensure $T(n,\si)\vert_{\si=0}=0$
in accord with the third thermodynamic law: $s(n,T)\vert_{T=0}=0$.

Altogether, one arrives at the sought energy function: 
\be\label{Efin}
e=m n  +Ce^{w\si}n^{\g},
\ee
with
\be
\g =1+w=\frac{c_p}{c_v}
\ee
and $c_p=1+ c_v$. 
Remind, that the specific heat $c_v$ is related
with the number $\nu$ of the internal degrees of freedom of the fluid
as $c_v=\nu/2$. In this, one has $\nu=3$ for the monatomic fluid at
any temperature, as
well as for the cold nonrelativistic two- and multiatomic fluids. For
the hot ultrarelativistic fluids, one has in the last two cases,
respectively, $\nu=5$  and $\nu=6$. Thus for the real fluids, one
has $ 1/3 \leq w\leq 2/3$. 
For the multiatomic fluids, the equations above are
be substituted by the picewise-smooth expressions due to the step-like
dependence~$c_v(T)$.

\paragraph{Radiation}

The radiation ($m=0$) can not be described consistently by the series
eq.~(\ref{Efin}). The reason is that the conserved current  $N^\mu$
and thus the continuity condition  are missing in this case.  To
balance between  the number of the equations of
motion and the number of the independent dynamical variables, 
there should be  left just one scalar variable. For consistency, one
chooses $s=n\si$, the Lagrangian being
in this case independent  of $n$ and  $\si$ apart. Now the  argument
about the choice of the Lagrangian set for eq.~(\ref{TLd}) does not
apply. Nevertheless, one can  retain the same
procedure treating the radiation as the marginal case with 
$E= E(|{\cal S|}/\sqrt {-g}\,)$, ${\cal S}^\mu$ being the entropy
current density.  From the requirement $T_m{}^\mu_\mu=0$,
there follows immediately the equation of state  $p = w\rho$, $w=1/3$.
With account for eq.~(\ref{rhop}) at $f=0$ and the relation
$n(\partial e/\partial n)_\si = s(\partial e/\partial s)_\si$ (prior
to cast  $\si$ aside),
one gets the energy function
\be\label{Erad}
e =Cs^\g,
\ee
with  $C$ being an integration constant and $\g=1+w=4/3$ ($\nu=6$,
$c_v=3$). The energy function $e$  and thus $\rho$ for
the radiation having been found, one can choose according to
eq.~(\ref{entropy}) the temperature  $T$ as the independent  variable
conjugate to~$s$:
\be
kT=\frac{\partial \rho}{\partial s}=(1+w)
C s^{w},
\ee
so that  $s\sim (k T)^3$, with $s(T)\vert_{T=0}=0$.
Thus, one gets in accord with the Stefan-Boltzman law: 
$\rho=3p\sim T^4$.

\paragraph{Vacuum}

In the same vein, considering the vacuum formally as the perfect fluid 
with the equation of state $p=-\rho$ ($m=0$, $\g=0$), one gets
$E=E_0$, the  constant independent of $n$ and~$\si$. Note that there
corresponds to the vacuum the specific heat $c_v=-1$  and thus,
formally, the number of the internal degrees of freedom  $\nu=-2$.

\section{Conclusion}

In conclusion, the classical theory of gravity with  the residual
UC is developed. In excess of the massless tensor graviton, the theory
describes the massive scalar one as a part of the metric.  Violating
the GC, the theory presents a viable alternative to the GR, as well as
to the generally covariant extensions thereof. The scalar graviton is
to be associated  naturally with (a part of) the DM. The
Lagrangian formalism for the  continuos medium interacting
with the scalar gravitons is developed. The effects due to the scalar
gravitons of possible interest in cosmology  are mainly twofold:
the  modification of the gravity and medium equations of motion,
the  distortion of the medium. These effects are typical for any
gravitational DM. 
The case study for the massive vector and scalar-tensor gravitons  is
expected to be given in the future.

\end{document}